\documentclass{elsart}
\usepackage{exscale}
\usepackage{amsmath}
\usepackage{amssymb}
\usepackage[dvips]{graphicx}
\usepackage{float}
\newcommand{\bea}{\begin{eqnarray}}
\newcommand{\eea}{\end{eqnarray}}

\newcommand{\nnl}{\nonumber\\}
\newcommand{\resig}{\mathrm{Re}\Sigma^\mathrm{ret}}
\newcommand{\ueff}{U_\mathrm{eff}}
\newcommand{\op}{(\omega,p)}
\newcommand{\sigmf}{\Sigma^\mathrm{mf}}
\newcommand{\mev}{\,\mathrm{MeV}}
\newcommand{\fm}{\,\mathrm{fm}}

\begin{document}

\begin{frontmatter}

\title{Short-Range Correlations  in Nuclear Matter at Finite Temperatures and High Densities\thanksref{sup}}
\thanks[sup]{Work supported by DFG and GSI.}
\author{F. Fr\"omel}\ead{Frank.Froemel@theo.physik.uni-giessen.de},
\author{H. Lenske},
\author{U. Mosel}
\address{Institut f\"ur Theoretische Physik, Universit\"at Giessen, Heinrich-Buff-Ring 16, D-35392 Giessen, Germany}

\begin{abstract}
The density and temperature dependence of the nucleonic single
particle spectral function in symmetric nuclear matter at finite
temperatures and densities beyond normal nuclear matter density is
investigated in a model emphasizing short-range correlations and
phase space aspects. A simple but self-consistent approach based
on quantum transport theory is used. In particular we consider the
density and temperature regime occuring e.g. during the core
collapse in a supernova explosion and the subsequent formation of
a neutron star. Mean-field effects are incorporated by a Skyrme
type potential.
\end{abstract}


\end{frontmatter}

\section{Introduction}

An interesting but only partially explored problem of nuclear
many-body theory are contributions from short-range correlations
to the binding properties of nuclear matter and finite nuclei. Of
particular interest for various regions of nuclear physics ranging
from heavy ion collisions to the physics of neutron star formation
is the density and temperature dependence of short-range
correlations in nuclear matter. In heavy ion collisions like those
planned at the Compressed Baryonic Matter (CBM) facility at GSI
densities of several times the nuclear matter saturation density
$\rho_0 = 0.16/\fm^3$ are reached. Such conditions are also
encountered in astrophysical scenarios. When a neutron star is
formed in a supernova explosion temperatures of several tens of
MeV and densities far beyond normal nuclear matter density arise.
The neutrinos emitted from the nascent neutron star are an
important signal that gives information about the ongoing cooling
processes. Nucleon correlations, however, affect directly the
neutrino opacity in dense nuclear matter. The neutrino cross
sections can be significantly reduced due to the presence of
correlations in the medium \cite{neutrino1,neutrino2}. Therefore
it is important to take nucleon correlations into account when
interpreting the neutrino spectra. 

So far there has been no
consistent investigation of the properties of the nucleon spectral
function that covers the full temperature and density range of
those scenarios. In this work we extend the approach to
short-range correlations in nuclear matter presented initially in
\cite{Lehr00,Lehr02} to the more general cases of arbitrary
temperatures and densities. Using direct relations between
correlation functions and collision integrals a simple but fully
self-consistent model for calculating the nucleon spectral
function in an iterative procedure was constructed in refs.
\cite{Lehr00,Lehr02}. The former calculations, however, have been
restricted to zero temperature and normal nuclear matter density.
The previously obtained results are in surprisingly good agreement
with much more sophisticated and numerically extremely involved,
"state-of-the-art" calculations utilizing the full machinery of
modern nuclear many-body theory \cite{ben,baldo}. In particular,
the significant population of high-momentum states in the nucleon
momentum distribution -- which is interpreted as an overall
measure for short-range correlations -- is described rather well.

In this paper, we reconsider this model and extend it to finite
temperatures and densities beyond the nuclear matter equilibrium
point. Different from \cite{Lehr00,Lehr02} this requires a
reliable description of the static mean-field self-energy which in
the former work could most conveniently be absorbed into a
redefinition of the chemical potential. In the present case an
explicit treatment of the density and momentum dependence of the
mean-field self-energy is necessary since the chemical potential
is temperature and density dependent. Calculations for the
equation of state for dense matter and finite temperatures have
been made using microscopic Hamiltonians, see e.g. \cite{wr,fp}.
In our model we incorporate the static properties of the
mean-field by an empirical, Skyrme-type energy density functional
\cite{scm,bf,vb}. By this choice thermodynamical consistency is
also ensured by fulfilling the Hugenholtz-van Hove theorem
\cite{HvH} at mean-field level. This approach gives rise to a potential
that is independent of energy but depends on momentum,
temperature, and density. However, because of the Skyrme-typical
restriction to terms of at most quadratic order in the single
particle momenta the momentum dependence can be re-expressed
immediately by an effective mass in the kinetic energy term.

Calculations similar to ours have been done by Alvarez-Ruso et al.
\cite{ruso} for densities from half to twice $\rho_0$ and
temperatures up to $20\mev$. In their "semi-phenomenological"
approach they have evaluated second order diagrams taking the
magnitude of the NN interaction from experiment. However, they did
not perform their calculations self-consistently and did not
consider long-range contributions from the mean-field. Note that
the basis of their work was also a model for cold nuclear matter
\cite{oset}. Self-consistent approaches to the nucleon spectral
function in nuclear matter have been studied theoretically before,
e.g. by Alm et al. \cite{alm} and in ref. \cite{djl}. More
recently, calculations using a self-consistent T-matrix approach
have been presented by Bo\.{z}ek \cite{boz1,boz2}. In \cite{boz1}
the temperature dependence of the spectral function was
investigated up to $T=20\mev$, but the density dependence has so
far not been investigated by Bo\.{z}ek.

In this work we go far beyond the temperature and density ranges
of \cite{ruso}-\cite{boz2}. To study the role of short-range
correlations in supernovae and heavy ion collisions at CBM we
consider temperatures up to $70\mev$ and densities from $\rho_0$
to $3\rho_0$. In section \ref{sec:model} we briefly outline our
model. The results of the calculations are presented in section
\ref{sec:results}. They are compared to other models (at $T=0$ and
finite temperatures) and we discuss the temperature and density
dependence of the short-range correlations in nuclear matter. A
summary is given in section \ref{sec:summary}.

\section{\label{sec:model}The Model}

The model we present here was introduced in \cite{Lehr00,Lehr02}
for cold ($T=0$) nuclear matter at saturation density $\rho_0$.
The underlying Green's function formalism is described in much
detail in \cite{kb,dan}. Thus we restrict ourselves here to a
short summary of the concept. We show in more detail how to modify
the original model \cite{Lehr00,Lehr02} for the application to
arbitrary densities and temperatures. Throughout this paper we
consider only symmetric nuclear matter.

On the two-particle--one-hole ($2p1h$) and one-particle--two-hole
($1p2h$) level, respectively, the collisional or polarization
self-energies $\Sigma^>$ and $\Sigma^<$ are given by: \bea
    \Sigma^\gtrless &=& g\int\frac{d^3p_2 d\omega_2}{(2\pi)^4}\frac{d^3p_3 d\omega_3}{(2\pi)^4}
                         \frac{d^3p_4 d\omega_4}{(2\pi)^4} (2\pi)^4 \delta^4(p+p_2-p_3-p_4)
                         \overline{|\mathcal{M}^2|}  \nnl
                      && \times g^\lessgtr(\omega_2,p_2) g^\gtrless(\omega_3,p_3) g^\gtrless(\omega_4,p_4), \label{eq:coll-se}
\eea
where $g=4$ is the nucleonic spin-isospin degeneracy factor
and $\overline{|\mathcal{M}|^2}$ is the square of the
nucleon-nucleon scattering amplitude averaged over spin and
isospin of the incoming and summed over spin and isospin of the
outgoing nucleons. We leave the structure of $\mathcal{M}$ open
for the moment and come back to it later. The correlation
functions $g^>$ and $g^<$ in eq.(\ref{eq:coll-se}) can be
rewritten in terms of the spectral function $a(\omega,p)$, \bea
    g^<\op&=& i a\op f\op, \nnl
    g^>\op&=& -i a\op (1- f\op), \label{eq:corr-spec}
\eea
with the phase space distribution function $f\op$. In thermal
equilibrium $f$ is just the Fermi distribution function depending
on the energy and, through the chemical potential $\omega_F$, on the density
of the system,
\bea
	f(\omega)=\frac{1}{1+e^{(\omega-\omega_F)/T}}.
\eea
The previous calculations were greatly simplified
by the fact that at zero temperature the Fermi distributions
reduce to step functions allowing for considerable simplifications
of eq.(\ref{eq:coll-se}) by taking advantage of closed form
analytic expressions \cite{Lehr00,Lehr02}.

The single particle spectral function in (\ref{eq:corr-spec}) is explicitly given as
\bea
    a\op=\frac{\Gamma\op}
            {(\omega-\frac{p^2}{2 m}-\sigmf-\resig\op)^2+\frac{1}{4}\Gamma^2\op},
            \label{eq:spectral}
\eea
where $\resig$ is the real part of the collisional
self-energy and $\Gamma$ is the width of the spectral function.
$\Gamma$ is directly related to the imaginary part of the retarded
self-energy,
\bea
    \Gamma\op=-2\textrm{Im}\Sigma^\mathrm{ret}\op
    =i(\Sigma^>\op-\Sigma^<\op).  \label{eq:gamma-def}
\eea
Note, that eq.(\ref{eq:spectral}) also includes the
mean-field self-energy $\sigmf=\sigmf(p,\rho,T)$ which is not
contained in $\Sigma^\gtrless$. The purely hermitian $\sigmf$
contains the contributions to the self-energy arising from ground
state tadpole diagrams describing mainly long-range correlations.

We can rewrite the first term in the denominator of
eq.(\ref{eq:spectral}) by expanding $\sigmf(p,\rho,T) \approx
\sigmf(k_F,\rho,T) + (p^2-k_F^2){\sigmf}'(k_F,\rho,T) + \ldots$ up
to $\mathcal{O}(p^2)$. The momentum dependent part of $\sigmf$ can
then be absorbed into the kinetic energy term by introducing a
density and temperature dependent effective mass
$\frac{1}{m^*}=\frac{1}{m}(1+2m{\sigmf}'),$ as successfully used,
e.g., in Skyrme density functionals \cite{scm,bf,vb}. We get:
\bea
\label{eq:mf_pot}
    \frac{p^2}{2 m}+\sigmf(p,\rho)+\resig\op
    & \cong &
    \frac{p^2}{2 m^*}+\ueff(\rho)+\resig\op ,
\eea
where the effective potential $\ueff$ is the momentum
independent part of $\sigmf$. We will discuss the role of $\ueff$
and $\resig$ in more detail below.

The equations (\ref{eq:coll-se})-(\ref{eq:gamma-def}) describe a
self-consistency problem. Typically, such problems cannot be
solved directly but iterative procedures are the appropriate
methods of solution. Hence, the spectral function is obtained
iteratively by the following procedure: Starting with an initial
guess for $a\op$ in the integrals of eq.(\ref{eq:coll-se}) an
approximation for the width $\Gamma$ is found. Inserting this
$\Gamma\op$ into eq.(\ref{eq:spectral}) yields an improved
expression for the spectral function that is inserted into
(\ref{eq:coll-se}) again. This procedure can be repeated until
convergence is achieved. In the language of Feynman diagrams the
solution of this self-consistency problem corresponds to the
non-perturbative summation over a whole class of diagrams by
reinserting the full in-medium Green's functions into the
self-energies.

As we have seen above there are two real contributions to the
self-energy. One is the mean-field self-energy that is purely real
and does not affect the width directly. The other contribution is
the real part of the collisional self-energy $\resig$. It is
related to the imaginary part of the polarization self-energy
(i.e., the width of the single particle spectral function) by a
dispersion relation, e.g. discussed in \cite{Lehr02}. There, the
importance of the dispersive contributions was investigated by
performing calculations with and without $\resig$. Although
analyticity is not conserved when $\resig$ is neglected
\cite{Lehr00} it was found that overall the results on spectral
functions did not change very much. Taking advantage of this
finding we will ignore the dispersive part for the present
investigations, thus simplifying the calculations considerably.
For a full treatment of $\resig$ we refer to \cite{Lehr00,Lehr02}.

More important for our present purpose of studying the density and
temperature dependence of short-range correlations are the
mean-field potential $\ueff=\ueff(\rho,T)$ and the effective mass 
$m^*=m^*(\rho)$. The net effect of the potential $\ueff$
is a temperature and density dependent energy shift of
the pole structure of the single particle propagators and hence
the spectral functions. For the investigations in
\cite{Lehr00,Lehr02} at $T=0$ and $\rho=\rho_0$ such a shift could
be absorbed into a re-definition of the chemical potential. This
is no longer possible when we look for the density and temperature
dependence of our results to which long-range correlations
contribute. 

The temperature and density dependence of the
single-particle potential in nuclear matter has been studied
before microscopically, ranging from self-consistent Brueckner and
Brueckner Hartree-Fock approaches e.g. in \cite{alm,djl} to the
variational calculations with 2- and 3-body forces by the Urbana
group \cite{wr,fp}. In a different context, the density and
momentum dependence of the variational mean-field was studied in
\cite{wr}. A re-analysis of these rather involved mean-field
results by a second order polynomial in $k^2-k^2_F$ leads to an
almost perfect description of the momentum dependence with an
effective mass of $m^*/m=0.69$. This confirms the effective mass
approach utilized in eq.(\ref{eq:mf_pot}). These findings also
confirm the use of a schematic description of the mean-field part,
in particular so because the static parts of the self-energies are
cancelled to a large extent in dynamical quantities like
propagators. Hence, we decide to describe the mean-field in a
Skyrme model \cite{scm,bf,vb} for the sake of a most transparent
and simple approach.

From the Skyrme energy density functional in standard notation
\cite{scm,bf,vb} $\ueff$ and $m^*$ 
in infinite symmetric nuclear matter are obtained by
\bea
    \ueff(\rho)&=&\frac{3}{4}t_0\rho+\frac{1}{16}(3t_1+5t_2)\tau+\frac{3}{16}t_3\rho^2, \\
    \frac{1}{2m^*}&=&\frac{1}{2m}+\frac{1}{16}(3t_1+5t_2)\rho,    
\eea
where  $t_0$, $t_1$, $t_2$, and $t_3$ are parameters of the Skyrme interaction \cite{scm,bf,vb}. The parameter set we have used here is given in Table \ref{tab:skyrme}.
$\tau$ is the (quasi-particle) kinetic energy density,
\bea
    \tau=\frac{2}{\pi^2}\int\limits_0^\infty dp ~ \frac{p^4}{e^{(p^2-k_F^2)/2m^*T}+1},
\eea
with $k_F^2=2m^*(\omega_F-\ueff)$.

It is well known that nuclear matter ceases to be bound beyond a
critical density $\rho_\mathrm{c}(T)$ where the repulsive kinetic
energy term overrules the nuclear attraction. For processes inside
a supernova this is, however, not a crucial problem. The core
collapse is not driven by nuclear, but by gravitational forces.
Eventually, the repulsive nuclear potential even stops the
collapse when several times nuclear matter density is reached.

For a complete definition of the approach we also have to specify
the scattering amplitude $\mathcal{M}$ in the collision integrals
(\ref{eq:coll-se}). Taking into account the observation of Lehr et
al. \cite{Lehr00,Lehr02} that the short-range correlations
primarily depend on global properties of the interactions on
ranges of the (repulsive) vector meson masses and the available
phase space we will also use a constant matrix
element in our calculations, corresponding to an effective zero-range 
contact interaction. 

The value of $\mathcal{M}$ will be chosen independent of temperature 
and density of the system. It is noteworthy that until now a 
description of short-range correlations including a fully 
self-consistent description of the interactions is still pending, 
although first attempts in that direction have been made. In such 
calculations, e.g. by the Rostock group \cite{alm} or, more 
recently, by Bozek \cite{boz1,boz2} significant variations of the 
interactions with temperature are not obtained. Accepting that 
short-range correlations are due to short-range interactions, a 
strong dependence of the interaction matrix element is not very 
likely, either.

In order to enforce the
convergence of integrations involving $\mathcal{M}$ and extending
over the space of $2p1h$ and $1p2h$ configurations, respectively,
we introduce a form factor such that the collision integrals in
eq.(\ref{eq:coll-se}) are cut off at large $\omega$. Lehr et al.
had introduced a form factor that acts on the total energies and
momenta of the collisions corresponding to eq.(\ref{eq:coll-se}).
This form factor did only affect the results at $\omega>\omega_F$.
For the occupied states below $\omega_F$ the results remained
unchanged. Due to the density dependence of the chemical
potential, however, this form factor is not a sensible choice for
our calculations.

Different to the approach in \cite{Lehr00,Lehr02} we use here a
form factor acting on the t-channel and suppressing processes with
large energy and momentum transfers (see \cite{feuster} for
details):
\bea
    F(\omega_\mathrm{t},\vec p_\mathrm{t})
    =\frac{\Lambda^4}{\Lambda^4+\left( m_\mu^2 -\omega_\mathrm{t}^2+p^2_\mathrm{t} \right)^2},
    \label{eq:ffactor}
\eea
where $\omega_\mathrm{t}=\omega-\omega_3$ and $\vec
p_\mathrm{t}=\vec p -\vec p_3$. The mass $m_\mu \sim
\mathcal{O}(m_\rho)$ corresponds to the typical mass of an
exchanged meson in nucleon scattering processes and $\Lambda$
defines the scale of the cutoff. Note that $F$ has no pole on the
real axis.

\section{\label{sec:results}Numerical Details and Results}

We have calculated the nucleon spectral function in nuclear matter
at temperatures from $10$ to $70\mev$ and at densities
$\rho=\rho_0 \cdots 3\rho_0$ ($\rho_0=0.16\fm^{-3}$). In the
calculations we used the Skyrme force type SIII' to generate a
mean-field potential and an effective mass $m^*$ (cf.
\cite{scm,bf,vb}). The corresponding set of parameters is given in
Table \ref{tab:skyrme}. In infinite nuclear matter at zero
temperature one finds for this choice of parameters a binding
energy of $-16.1\mev$ and a mass ratio of $m^*/m=0.76$ at a Fermi
momentum of $k_F=1.29\fm^{-1}$ ($\rho=\rho_0$). For the mean-field
potential a value of $\ueff=-61.6\mev$ is found. Compared to the
variational results of Wiringa \cite{wr} the saturation properties
are slightly different with an equilibrium point shifted to a
somewhat lower density. The larger effective mass indicates a
weaker momentum dependence of the SIII' interaction.

\begin{figure}
    \begin{center}\includegraphics[scale=1]{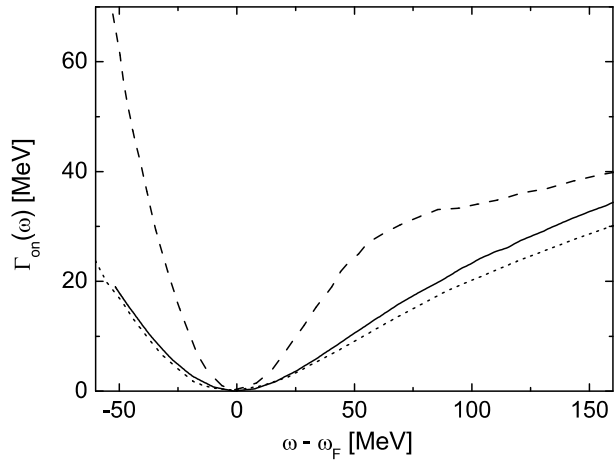}\end{center}
    \caption{\label{fig:os-comp}The on-shell width $\Gamma_\mathrm{on}(\omega)$ as a function of $\omega-\omega_F$ for $T=0$ and normal nuclear matter density. The solid line shows the result of our calculation, the dashed line the result of Benhar et al. \cite{ben}, and the dotted line the result of Baldo et al. \cite{baldo}.}
\end{figure}

The averaged scattering amplitude was adjusted to the
value\footnote{This value differs from the one used in
\cite{Lehr00,Lehr02} by the factor $m/m^*\approx 1.4$.}
$(\overline{|\mathcal{M}|^2})^{1/2}=309 \mev \fm^3$  so that the
results of the many-body calculations of Benhar et al. for $T=0$
\cite{ben} are reproduced. The cutoff parameter $\Lambda$ of the
form factor (\ref{eq:ffactor}) was set to $1.2\,\mathrm{GeV}$ and
the mass $m_\mu$ to $600\mev$. To verify that this is a reasonable
choice for the parameters we compare the on-shell width
$\Gamma_\mathrm{on}(\omega)=\Gamma(\omega,p(\omega))$, where
$p(\omega) = (2m^*(\omega-\ueff))^{1/2}$, from our calculation to
the results of Benhar et al. \cite{ben} and Baldo et al.
\cite{baldo}. In Fig. \ref{fig:os-comp} the on-shell widths are
shown for normal nuclear matter density and $T=0$. As it can be
seen, our result is in good agreement with the sophisticated
many-body calculation of Benhar et al. and supports the choice of
the parameters. Note that the discrepancies between the results of
Baldo and Benhar are explained in \cite{ben} by the different NN
interactions that are employed.

\begin{figure}
    \begin{center}\includegraphics[scale=1]{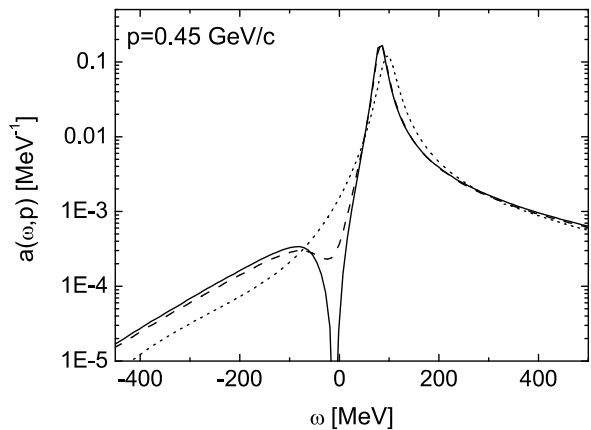}\end{center}
    \caption{\label{fig:spec-temp}The nucleon spectral function at $T=0\mev$ (solid line), $T=10\mev$ (dashed line) and $T=50\mev$ (dotted line) for normal nuclear matter density. All cuts were made at a constant momentum of $p=0.45\,\mathrm{GeV}$.}
\end{figure}

In the following we will show the results of our calculations and
investigate the density and temperature dependence of the
short-range correlations. First, we present cuts of the nucleon
spectral function at a constant momentum to illustrate the general
role of temperature and density. Fig. \ref{fig:spec-temp} shows
the spectral function for temperatures of $0$, $10$ and $50\mev$
at normal nuclear matter density. The effect of higher
temperatures is clearly visible. The gap at the Fermi energy gets
smeared out for $T=10\mev$ and has almost vanished at $T=50\mev$.
This phenomenon is directly related to the fact that for
non-vanishing T -- in contrast to a system at zero temperature --
states at energies above $\omega_F$ are populated. The states at
the Fermi surface cease to be quasi-stable.

\begin{figure}
    \begin{center}\includegraphics[scale=1]{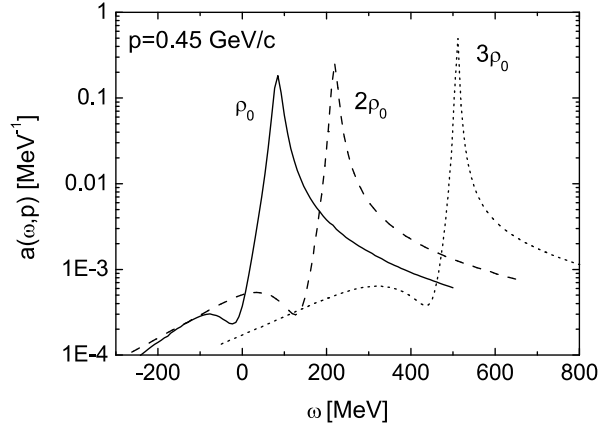}\end{center}
    \caption{\label{fig:spec-rho}The nucleon spectral function for normal nuclear matter density (solid line), a two times higher density (dashed line), and a three times higher density (dotted line) at $T=10\mev$. All cuts were made at a constant momentum of $p=0.45\,\mathrm{GeV}$.}
\end{figure}

In Fig. \ref{fig:spec-rho} the spectral function is shown for
three different densities at a constant temperature of $10\mev$.
Since the chemical potential depends on the density the three
curves are shifted along the energy axis. Note that the relation
$\mu \sim \rho^{1/3}$ for the free Fermi gas does not hold here as
it can be directly seen for $T=0$, where
$\mu=\tau+\mathrm{Re}(\Sigma(k_F))$. One should not get mislead by
the observation that the on-shell peaks at lower densities are
broader than the peak at the highest density. This effect is due
to the fact that the on-shell energy lies closer to the Fermi
energy for the highest density. This induces a stronger
suppression of the on-shell peak than in the other cases.

\begin{figure}
    \begin{center}\includegraphics[scale=1]{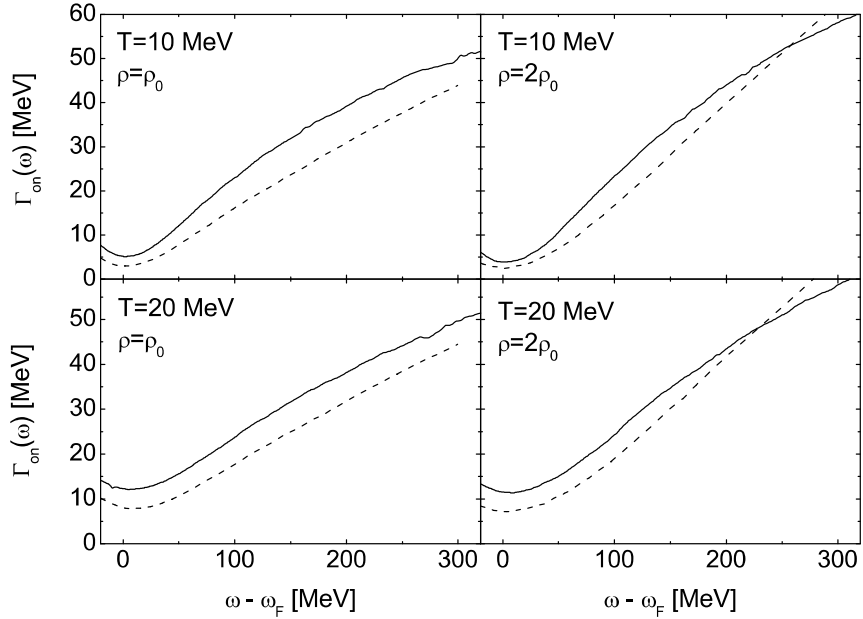}\end{center}
    \caption{\label{fig:os-luis}The on-shell width $\Gamma_\mathrm{on}(\omega)$ as a function of $\omega-\omega_F$ for two temperatures and densities. The solid lines show the results of our calculations, the dashed lines show the results of Alvarez-Ruso et al. \cite{ruso}.}
\end{figure}

Figs. \ref{fig:spec-temp} and \ref{fig:spec-rho} give a first
impression of the spectral functions. To investigate the
temperature and density dependence of the short-range
correlations, however, we use the better suited on-shell width of
the spectral function, $\Gamma_\mathrm{on}(\omega)$. In Fig.
\ref{fig:os-luis} we show our results for
$\Gamma_\mathrm{on}(\omega)$ at temperatures of $10$ and $20\mev$
and densities of $\rho_0$ and $2\,\rho_0$. Also shown are the
results of Alvarez-Ruso et al. \cite{ruso}. As we have already
seen in Figs. \ref{fig:spec-temp} and \ref{fig:spec-rho} the width
of the states at the Fermi surface does not drop to zero for
finite temperatures. Increasing the temperature leads to larger
values of $\Gamma_\mathrm{on}$ in the vicinity of the Fermi
surface while the size and the shape of $\Gamma_\mathrm{on}$ do
not change significantly at higher energies. Increasing the
density shows the opposite effect. $\Gamma_\mathrm{on}$ is left
unchanged in the vicinity of the Fermi surface while the slope at
high energies increases. Compared to \cite{ruso} our results seem
to be shifted by approximately $5\mev$ to higher values of
$\Gamma_\mathrm{on}$. The shift can be explained by the
differences in the NN interactions that were used in both
calculations. The shape of the curves is in good agreement for
$\omega-\omega_F < 150\mev$. At higher energies the form factor
that was used in our calculations affects the shape of
$\Gamma_\mathrm{on}$, e.g. leading to the observed curvature in
our results.

Qualitatively our results (for $\rho=\rho_0$ and $T\le20\mev$) are
similar to those obtained in the self-consistent T-matrix approach
by Bo\.{z}ek \cite{boz1,boz2} but on a quantitative level
significant differences are found. However, the deviations range
within the uncertainties introduced using different
nucleon-nucleon interactions in fully microscopic calculations.
This is evident by comparing the results in \cite{boz2},
based on the Paris NN potential, e.g. to the variational results
of Benhar et al. \cite{ben}, obtained with the Argonne NN
potential. The reason for these deviations is found in the fact
that by the way of deriving these empirical interactions from NN
data they are guaranteed to agree in their predictions of on-shell
quantities, as e.g. free space NN scattering phase shifts, but
might differ in the off-shell region. While we work with a
constant coupling in combination with a t-channel form factor
Bo\.{z}ek tends to use strongly momentum dependent interactions
like the Paris potential (see \cite{boz2} for details). Similar
observations are discussed in \cite{boz2} when results of
different approaches -- including \cite{Lehr00} and \cite{boz1} --
are compared. Note that the same effect has already shown up in
Fig. \ref{fig:os-comp} for the results of Benhar \cite{ben} and
Baldo \cite{baldo}.

\begin{figure}
    \begin{center}\includegraphics[scale=1]{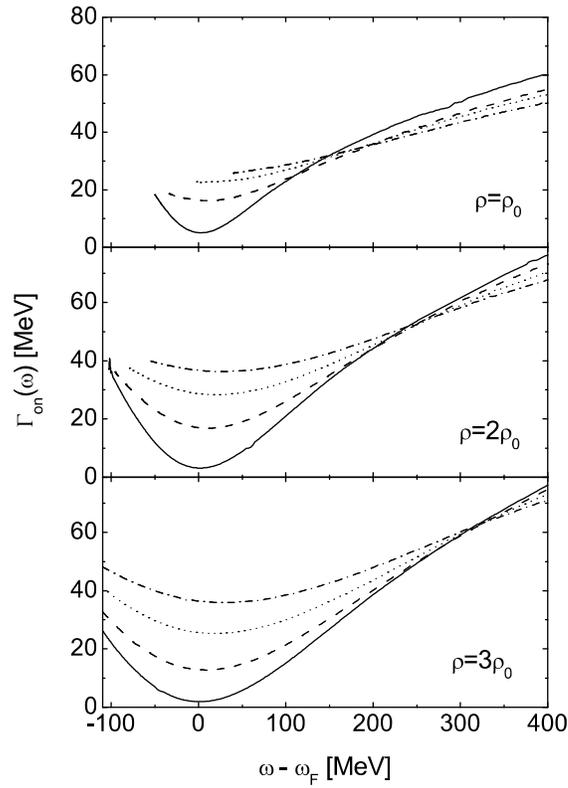}\end{center}
    \caption{\label{fig:spec-cuts}The on-shell width $\Gamma_\mathrm{on}(\omega)$ as a function of $\omega-\omega_F$ for different temperatures and densities.
    Results for $T=10,30,50,70\mev$ are indicated by solid, dashed, dotted and dashed-dotted lines, respectively.}
\end{figure}

The temperature dependence of the on-shell width at various
densities is illustrated in Fig. \ref{fig:spec-cuts}. The effect of
filling the gap at the Fermi surface with increasing temperature
is clearly visible. At all densities one observes a levelling off
in the energy dependence of $\Gamma_\mathrm{on}(\omega)$ at the highest
temperatures indicating the gradual changes in the available phase
space in a heated system.

\begin{figure}
    \begin{center}\includegraphics[scale=1]{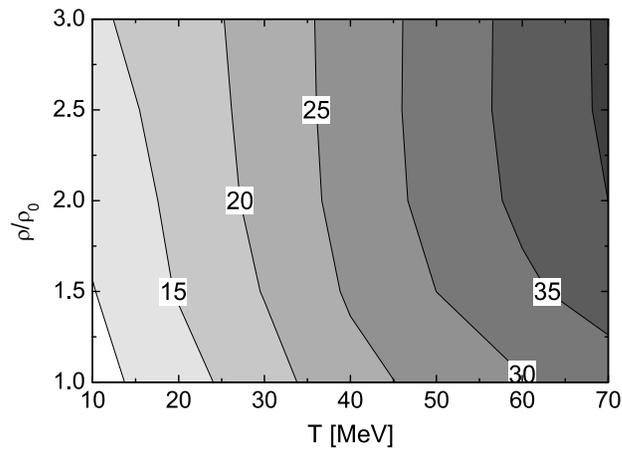}\end{center}
    \caption{\label{fig:contour} The on-shell width $\langle\Gamma_\mathrm{on}\rangle$ in the $(T,\rho)$-plane ($\Gamma$ in MeV). For each temperature and density $\Gamma_\mathrm{on}$ has been averaged over the occupied states.}
\end{figure}

The behavior seen in Fig. \ref{fig:spec-cuts} deserves a closer
inspection. Before doing so we should take into account that the
chemical potential $\omega_F$ depends strongly on the nucleon
density, as was shown in Fig. \ref{fig:spec-rho}. Thus one has to
be careful at which energy $\omega$ and momentum $p(\omega)$,
respectively, the widths for different temperatures and densities
are compared. This problem is avoided by averaging the on-shell
width over the occupied states and compare this
$\langle\Gamma_\mathrm{on}\rangle$ for the different $\rho$ and $T$. Fig.
\ref{fig:contour} shows the overall behavior of
$\langle\Gamma_\mathrm{on}\rangle$ for the full temperature and density range
that was covered in our calculations. Note that one gets a very
similar picture when $\Gamma_\mathrm{on}(\langle\omega\rangle)$, i.e. the
on-shell width at the averaged energy of the occupied states, is
displayed instead. The general features of this figure are thus
universal. It can be seen in Fig. \ref{fig:contour} that the
on-shell width ranges from $10\mev$ at low temperatures up to
$40\mev$ at high temperatures. $\langle\Gamma_\mathrm{on}\rangle$ rises almost
linearly with the temperature at medium and high densities. At
normal nuclear matter density the dependence is only linear for
temperatures below $\approx 45\mev$ and gets weaker at higher
temperatures. The density dependence of $\langle\Gamma_\mathrm{on}\rangle$
that can be derived from Fig. \ref{fig:contour} has a more
complicated structure. As long as the density is low
($\rho<2\rho_0$) $\langle\Gamma_\mathrm{on}\rangle$ does increase with the
density just as one would expect naively. At densities above
$2\rho_0$, however, the density dependence becomes very weak -- in
particular for temperatures above $30$ - $40\mev$. Since we take
the on-shell width as a measure for short-range correlations this
means that the correlations saturate at densities above two times
normal nuclear matter density.

The explanation for the saturation is rather simple.
Pauli-blocking suppresses nucleon collisions (the source of the
correlations) with final states below the Fermi surface. Since the
chemical potential depends strongly on the nucleon density more
(kinematically allowed) final states are blocked at higher $\rho$.
At high $\rho$ there are also more collision partners (initial
states) for a particle with given energy and momentum but this
cannot compensate the suppression from Pauli-blocking. To
understand this one has to look at the total collision rate for a
nucleon with given $\omega$ and $p$ with the nucleons of the
medium (cf. eq.(\ref{eq:coll-se})). There are two final states
that are affected by Pauli-blocking but there is only one initial
state from below the Fermi surface. Thus the effect of increased
Pauli-blocking is more important than the gain of initial states.

\begin{figure}
    \begin{center}\includegraphics[scale=1]{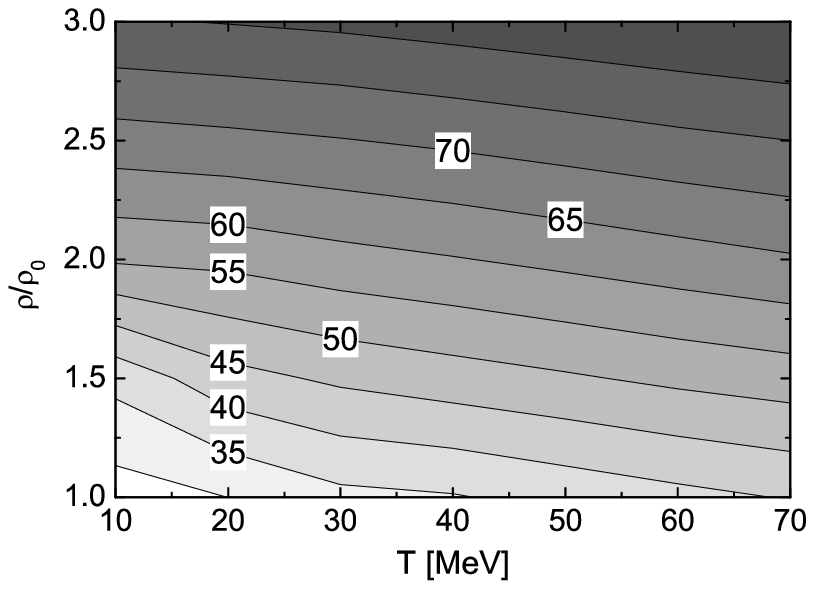}\end{center}
    \caption{\label{fig:contour_np} The on-shell width $\langle\Gamma_\mathrm{on}\rangle$ in the $(T,\rho)$-plane ($\Gamma$ in MeV) when Pauli-blocking is switched off in the calculations. For each temperature and density $\Gamma_\mathrm{on}$ has been averaged over the occupied states.}
\end{figure}

To verify that this explanation is correct we have made additional
calculations in which Pauli-blocking was switched off. The result
is shown in Fig. \ref{fig:contour_np}. It can be clearly seen that
without Pauli-blocking there is a linear dependence of
$\langle\Gamma_\mathrm{on}\rangle$ on both, temperature and density, over the
full range covered by our calculations. This is exactly the
picture one would naively expect when Pauli-blocking is not taken
into account. Thus Pauli-blocking is in fact the correct
explanation for the saturation of the short-range correlations at
high densities.

\section{\label{sec:summary}Summary and Conclusions}

In this paper we have presented an approach to short-range
correlations in nuclear matter at finite temperatures and high
densities. Based on the work of Lehr et al. \cite{Lehr00,Lehr02}
we have constructed a simple but self-consistent model that goes
beyond the quasi-particle approximation. The interactions between
the nucleons were described by a constant matrix element in
combination with a form factor that limits the energy and momentum
transfer in the collisions. In \cite{Lehr00,Lehr02} the averaged
coupling constant has led to excellent results for a system at
zero temperature.  We have not calculated the mean-field
contributions self-consistently within the model. Instead, a
Skyrme-type interaction model was incorporated to assure a
realistic thermodynamical behavior of the chemical potential.

We have calculated the spectral function of nucleons in symmetric
nuclear matter at temperatures from $10$ to $70\mev$ and densities
from $\rho_0$ to $3\rho_0$. This has been the first time that this
large regime is investigated consistently within the same model.
It was shown that our results are in good agreement with other
calculations at temperatures up to $20\mev$ and densities up to
$2\,\rho_0$. The width of the spectral function was used to get
informations about the temperature and energy dependence of the
short-range correlations. We have found that the correlations
scale approximately linear with temperature. At high densities,
however, the correlations saturate, an effect that can be
explained by Pauli-blocking.

The model we have presented here is open for further improvements.
Using separate spectral functions for protons and neutrons and
introducing an isospin dependent coupling it would also be
possible to investigate isospin asymmetric nuclear matter.

\section*{Acknowledgments}

We thank Stefan Leupold and J\"urgen Lehr for helpful discussions
during the preparation of this work. In addition we would like to
thank J. Lehr for providing the original computer code that was
used in \cite{Lehr00}.

\newpage

\begin{table}
    \begin{center}
      \small
        \begin{tabular}{|c|c|c|c|c|c|} \hline
            \small$t_0$ &   \small$t_1$ & \small$t_2$ & \small$t_3$ & \small$x_0$ & \small$W_0$ \\
            \small $(\mathrm{MeV}\cdot\mathrm{fm}^3)$ & \small $(\mathrm{MeV}\cdot\mathrm{fm}^5)$ &
            \small $(\mathrm{MeV}\cdot\mathrm{fm}^5)$ & \small $(\mathrm{MeV}\cdot\mathrm{fm}^6)$
            & &
            \small$(\mathrm{MeV}\cdot\mathrm{fm}^5)$  \\ \hline \hline
            \small -1133.4 & \small 395 & \small -95 & \small 14000 & \small 0.49 & \small 120 \\ \hline
        \end{tabular}
        \normalsize
    \end{center}
    \caption{\label{tab:skyrme}The parameters of the Skyrme force SIII' \protect\cite{scm} used in the calculations. For completeness this table also contains the spin-interaction para\-meter $x_0$ and the spin-orbit parameter $W_0$.}
\end{table}

\newpage

\newpage

\begin{thebibliography}{99}

\bibitem{neutrino1}R. F. Sawyer, Phys. Rev. Lett. \textbf{75} (1995) 2260.

\bibitem{neutrino2} S. Reddy, J. Pons, M. Prakash, J.M. Lattimer, astro-ph/9802312.

\bibitem{Lehr00}
J. Lehr, M. Effenberger, H. Lenske, S. Leupold, U. Mosel,
Phys. Lett. \textbf{B483} (2000) 324.

\bibitem{Lehr02}J. Lehr, H. Lenske, S. Leupold, U. Mosel,
Nucl. Phys. \textbf{A703} (2002) 393.

\bibitem{ben} O. Benhar, A. Fabrocini, S. Fantoni, Nucl. Phys. \textbf{A505} (1989) 267;
O. Benhar, A. Fabrocini, S. Fantoni,  Nucl. Phys. \textbf{A550} (1992) 201.

\bibitem{baldo} M. Baldo, I. Bombaci, G. Giansiracusa, U. Lombardo, C. Mahaux, R. Sartor,
Nucl. Phys. \textbf{A545} (1992) 741.

\bibitem{wr} R. B. Wiringa,
Phys. Rev. \textbf{C38} (1988) 2967.

\bibitem{fp} B. Friedman, V. R. Pandharipande,
Nucl. Phys. \textbf{A361} (1981) 592.

\bibitem{scm}G. Sauer, H. Chandra, U. Mosel,
Nucl. Phys. \textbf{A264} (1976) 221.

\bibitem{bf}M. Beiner, H. Flocard, N. van Giai, P. Quentin,
Nucl. Phys. \textbf{A238} (1975) 29.

\bibitem{vb}D. Vautherin, D. M. Brink,
Phys. Rev. \textbf{C5} (1972) 626.

\bibitem{HvH}
N.M. Hugenholtz, L. van Hove, {\em Physica} {\bf 24} 363 (1958).

\bibitem{ruso} L. Alvarez-Ruso, P. Fernandez de Cordoba, E. Oset,
Nucl. Phys. \textbf{A606} (1996) 407.

\bibitem{oset} P. Fernandez de Cordoba, E. Oset,
Phys. Rev. \textbf{C46} (1992) 1697.

\bibitem{alm} T. Alm, G. R\"opke, A. Schnell, N. H. Wong, H. S. K\"ohler,
Phys. Rev. \textbf{C53} (1996) 2181.

\bibitem{djl} F. de Jong, H. Lenske,
Phys. Rev. \textbf{C56} (1997) 154.

\bibitem{boz1} P. Bo\.{z}ek, Phys. Rev. \textbf{C59} (1999) 2616.

\bibitem{boz2} P. Bo\.{z}ek, Phys. Rev. \textbf{C65} (2002) 054306.

\bibitem{kb} L. P. Kadanoff, G. Baym,
\textit{Quantum Statistical Mechanics,}
Benjamin, New York (1962).

\bibitem{dan} P. Danielewicz,
Ann. Phys. \textbf{152} (1984) 305.

\bibitem{feuster} T. Feuster, U. Mosel,
Phys. Rev. \textbf{C58} (1998) 457.

\end{thebibliography}
\end{document}